\documentclass[12pt,oneside]{article}
\usepackage{epsfig, cite}
\usepackage{color}
\usepackage{ifthen}
\usepackage{graphicx}

\def\p{\partial}

\def\half{{1\over 2}}
\def\({\left(}
\def\){\right)}
\def\[{\left[}
\def\]{\right]}

\def\e{\begin{equation}}
\def\q{\end{equation}}
\def\m{\begin{eqnarray}}
\def\n{\end{eqnarray}}

\begin{document}
\thispagestyle{empty} \setcounter{page}{0}

\vspace{2cm}

\begin{center}
{\Large Gravitational Correction and Weak Gravity Conjecture}

\vspace{1.4cm}

Qing-Guo Huang

\vspace{.2cm}

{\em School of physics, Korea Institute for
Advanced Study,} \\
{\em 207-43, Cheongryangri-Dong,
Dongdaemun-Gu, } \\
{\em Seoul 130-722, Korea}\\
\end{center}

\vspace{-.1cm}

\centerline{{\tt huangqg@kias.re.kr}} \vspace{1cm}
\centerline{ABSTRACT}
\begin{quote}
\vspace{.5cm}

We consider the gravitational correction to the running of gauge
coupling. Weak gravity conjecture implies that the gauge theories
break down when the gravitational correction becomes greater than
the contribution from gauge theories. This observation can be
generalized to non-Abelian gauge theories in diverse dimensions and
the cases with large extra dimensions.

\end{quote}
\baselineskip18pt

\noindent

\vspace{5mm}

\newpage

\setcounter{equation}{0}

In \cite{Vafa:2005ui}, Vafa suggested that gravity and the other
gauge forces can not be treated independently and the vast series of
semi-classically consistent effective field theories which belong to
swampland are actually inconsistent after gravity is included. The
authors in \cite{Arkani-Hamed:2006dz} proposed the weak gravity
conjecture which can be most simply stated as gravity is the weakest
force. This conjecture implies that in a four-dimensional theory
with gravity and a U(1) gauge theory, there is a new intrinsic UV
cutoff for the U(1) gauge theory \e \Lambda\sim gM_4 \label{lgm}, \q
where $M_4=1.4\times 10^{19}$ GeV is the four-dimensional Planck
scale and $g$ is the U(1) gauge coupling.

In \cite{Huang:2006hc}, some heuristic arguments on the weak gravity
conjecture in the asymptotical dS and AdS space implies that a lower
bound on the U(1) gauge coupling $g$, or equivalently, the absolute
value of the cosmological constant gets an upper bound \e |\rho_V
|\leq g^2 M_4^4, \q in order that the U(1) gauge theory can survive
in four dimensions. This result has a simple explanation in string
theory, i.e. the string scale $\sqrt{\alpha '}$ should not be
greater than the size of the cosmic horizon. On the other hand, the
most exciting possibility raised by large extra dimensions
\cite{Arkani-Hamed:1998rs,Antoniadis:1998ig,Arkani-Hamed:1998nn,Cremades:2002dh,Kokorelis:2002qi,Floratos:2006hs}
is that the fundamental Planck scale may be much lower than the
apparent four-dimensional Planck scale. This implies that we may
begin to experimentally access the dynamics of quantum gravity
sooner than previously anticipated. In \cite{Huang:2006pn}, we
proposed that the intrinsic UV cutoff for U(1) gauge theory with
large extra dimensions is proportional to the fundamental Planck
scale, not the four-dimensional Planck scale. This new energy scale
predicted by weak gravity conjecture may be relevant to the physics
at LHC. In \cite{Banks:2006mm}, Banks et al. generalized the weak
gravity conjecture to higher dimensions. Some other related topics
are discussed in
\cite{Huang:2006tz,Kachru:2006em,Li:2006jj,Adams:2006sv,Li:2006vc,Ooguri:2006in,Kats:2006xp,Medved:2006ht}.

In this paper, we investigate the gravitational correction to the
Callan-Symanzik $\beta$ function in diverse dimensions and the case
with large extra dimensions, and provide a new viewpoint on the weak
gravity conjecture. We do not take into account three or lower
dimensions as gravity does not contain propagating degree of freedom
in these dimensions, even though some of our results may be
applicable to three dimensional cases as well.

Robinson and Wilczek in \cite{Robinson:2005fj} consider the one-loop
gravitational correction to the running of gauge theory couplings in
four dimensions. See other consideration of the correction to the
$\beta$ function: for example \cite{Hossenfelder:2004up}. They take
into account the Feynman diagrams involving a vertex dressed by
graviton exchange. The gluon-graviton vertex in four dimensions is
proportional to $\Lambda/M_4$, where $\Lambda$ is the energy scale.
After considering the graviton exchange, one-loop $\beta$ function
for the running of gauge coupling takes the form \e
\beta\equiv{dg\over d\ln\Lambda}=-{b_0\over
(4\pi)^2}g^3-a_0\({\Lambda\over M_4}\)^2g. \label{bf} \q The first
term on the right hand side of (\ref{bf}) takes the form of familiar
non-gravitational contribution and the second term includes the
gravitational contribution. Because gravitons don't carry any gauge
charges, the value of $b_0$ is determined by the matter content and
the gauge group. The unknown coefficient $a_0$ in eq. (\ref{bf}) is
determined to be $a_0={3/ \pi}$ in \cite{Robinson:2005fj}. When
gravitational effects become significant, the local effective gauge
theory is broken down. Requiring that the gravitational correction
is not greater than the contribution from gauge theory yields \e
\Lambda\leq CgM_4, \label{gcc} \q where \e C={1\over
4\pi}\sqrt{|b_0|\over a_0}.\q This result is just the same as
\cite{Arkani-Hamed:2006dz}. Now weak gravity conjecture is
interpreted as the condition for that the gravitational correction
is smaller than the contribution from gauge fields. The above
argument can be easily generalized to non-Abelian gauge field
theories. For example, $b_0={11\over 3}N$ for pure SU(N) Yang-Mills
theory. The weak gravity conjecture predicts \e \Lambda \leq C_N
\sqrt{Ng^2} M_4, \q where $C_N\simeq 0.16$. The combination with
$Ng^2$ is nothing but the 't Hooft coupling. It is just what we
expect.

Here a suggestive coefficient $C_1$ is obtained. In standard model,
$b_0=-41/6$ for U(1)$_Y$ and thus $C=0.21$. For SU(2)$_L$,
$b_0=19/6$ and $C=0.14$; for SU(3)$_C$, $b_0=7$ and $C=0.22$.
However, it is well known that the pure U(1) electromagnetism and
${\cal N}=4$ super-Yang-Mills in four dimensions are scale free
theories with $b_0=0$. Weak gravity conjecture implies these two
theories cannot self-consistently survive with gravity.

In the case with large extra dimensions, the fundamental Planck
scale can be much lower than the apparent four-dimensional Planck
scale. If the fundamental Planck scale is roughly 1 TeV, it will
open a window to detect quantum gravity in LHC in the near future.
Since the fundamental Planck is quite low, we expect that gravity
provides a significant correction on the one-loop $\beta$ function.
In this scenario, gravitons propagate in the whole spacetime, but
the gauge fields only propagate in four dimensions. The interaction
between graviton and gauge fields is described by the action \e S=
{1\over 2\kappa_d^2}\int d^dx\sqrt{-det (g_{mn})}R+\int d^4x
\sqrt{-det (g_{\mu\nu})} {1\over 4} \hbox{Tr}(F^2), \label{ac}\q
where $\kappa_d^2\sim G_d$ is the Newton coupling constant in $d$
dimensions. Consider the quantum fluctuation of the gravitational
degrees of freedom around the Minkowski metric as \e
g_{mn}=\eta_{mn}+\kappa_d h_{mn}. \q The action (\ref{ac}) becomes
\e S\sim \int d^dx (\p h)^2 + \int d^4x \kappa_d \eta^{\mu\rho}
\p_\mu A_\nu \p_\rho A_\sigma h^{\nu\sigma}+\cdot \cdot \cdot.\q The
interaction term between graviton and gauge field is proportional to
positive powers of $\kappa_d$. Dimensional analysis implies that the
gluon-graviton vertex is proportional to $\kappa_d \Lambda^{d-2\over
2}$. If we take the low energy limit, all interaction terms between
gauge field and graviton drop out. The running of gauge coupling is
governed by a modified $\beta$ function which is given by \e
\beta=-{b_0\over (4\pi)^2}g^3-c_0\kappa_d^2\Lambda^{d-2}g=-{b_0\over
(4\pi)^2}g^3-c_0\({\Lambda\over M_d}\)^{d-2}g, \label{bec} \q where
$M_d=G_d^{-{1\over d-2}}$ is the $d$-dimensional Planck scale.
Demanding that the gravitational correction is less than the gauge
contribution leads to \e \Lambda\leq C_E g^{2\over d-2}M_d,
\label{wex}\q with \e C_E=\({1\over 16\pi^2}\left|{b_0\over
c_0}\right|\)^{1\over d-2}. \q Eq. (\ref{wex}) is the same as the
result in \cite{Huang:2006pn}. Similarly, for the SU(N) gauge
theory, $b_0\sim N$ and then $\Lambda \leq (Ng^2)^{1\over d-2}M_d$.
In \cite{Gogoladze:2006pa}, the authors use the results of Robinson
and Wilczek as a testing ground to probe where the physical
gravitational scale may be. But for the case with large extra
dimensions, we should use eq. (\ref{bec}).

In the scenario with large extra dimensions, not only the black
holes are possibly produced \cite{Giddings:2001bu,Alberghi:2006km},
but also a new intrinsic UV cutoff for gauge theories in standard
model emerges. We can also detect the gravitational effects
according to the modification for the running of the gauge coupling.
The value of unknown coefficient $c_0$ will be determined in the
future.

Our previous arguments on the gauge theories coupled to gravity can
be generalized to the cases in higher dimensions. In $d$ dimensions,
gauge coupling has dimensions of $[mass]^{2-d/2}$. For $d>4$, gauge
coupling takes dimensions of positive power of length and gauge
interaction becomes irrelevant. The gauge coupling depends on the
energy scale $\Lambda$ according to \e {dg_d\over d\ln
\Lambda}=-f_0\Lambda^{d-4}g_d^3-h_0G_d\Lambda^{d-2}g_d,
\label{dgc}\q where $g_d$ is $d$-dimensional gauge coupling, $f_0$
and $h_0$ are the numerical coefficients. The factors of $\Lambda$
in (\ref{dgc}) have been inserted by dimensional analysis. Again
requiring that the gravitational correction is less than the
contribution from gauge fields yields \e \Lambda\leq
g_dM_d^{d-2\over 2}, \q which is just the same as that in
\cite{Banks:2006mm}. For SU(N) gauge theories, we just need to
replace $g$ with $\sqrt{N}g$.

We are also interested in investigating the weak gravity conjecture
in the asymptotical de Sitter and anti-de Sitter space. Following
the idea of \cite{Huang:2006hc}, these effective field theories
breaks down when the curvature radius of the background is less than
the shortest reliable distance for the field theories $1/\Lambda$.
The curvature radius of the background is roughly $L\sim
1/\sqrt{G_d|\rho_V|}$, where $\rho_V$ is the cosmological constant.
Requiring $L>1/\Lambda$ yields an upper bound on the cosmological
constant $\rho_V$ in $d$ dimensions \e |\rho_V|\leq
Ng_d^2M_d^{2(d-2)}  \label{cc}\q for SU(N) gauge theory. For U(1)
gauge theory in four dimensions, the result is just the same as that
in \cite{Huang:2006hc}.

Consider the brane world scenario on Dp-brane. Now $d=p+1$. The
gauge coupling for U(1) gauge theory on a Dp-brane is \e g_d\sim
g_s^{\half}M_s^{-{p-3\over 2}}.\q Taking toroidal compactification,
the $d$-dimensional Planck scale is related to the size of the extra
dimensions $R$ by \e M_d^{p-1}\sim g_s^{-2}M_s^{p-1}(M_sR)^{9-p}.\q
The tension of Dp-brane \e T_p\sim {M_s^{p+1}\over g_s}\q provides
an effective cosmological constant on the brane. Eq. (\ref{cc}) is
translated into \e g_s\leq (M_sR)^{(9-p)}.\label{gmr}\q On the other
hand, the Hubble constant on the brane takes the form \e H^2\sim
{T_p\over M_d^{p-1}}\sim {g_sM_s^2\over (M_sR)^{9-p}}.\q With the
viewpoint of string theory, requiring that the string length should
be shorter than the curvature radius of the background, $l_s\leq
H^{-1}$, yields $g_s\leq (M_sR)^{(9-p)}$ which is the same as
(\ref{gmr}); otherwise, string can be not taken as a point-like
particle and the full string theory should be involved.

Similarly, for a stack of $N$ Dp-branes, the gauge group becomes
SU(N) for the open string modes on these D-branes and the effective
cosmological constant is given by $\rho_V\sim NT_p$. Now eq.
(\ref{cc}) becomes \e g_s\leq (M_sR)^{9-p}. \label{dbc}\q On the
other hand, the Hubble constant on these branes takes the form \e
H^2\sim {NT_p\over M_d^{p-1}}\sim {Ng_sM_s^2\over (M_sR)^{9-p}}.\q
Requiring $l_s\leq H^{-1}$ leads to \e g_sN\leq (M_sR)^{(9-p)},\q
which is more stringent than (\ref{dbc}). It seems that our
constraint on the SU(N) gauge theory in the asymptotical de Sitter
space is quite loose.

In this paper, we provide a new observation on the weak gravity
conjecture which is independent on the arguments based on black hole
\cite{Arkani-Hamed:2006dz,Banks:2006mm}. The gravitational
correction should be less than the contribution from the gauge
theories; otherwise, the effective description of the gauge theories
breaks down. The straightforward framework for quantum gravity,
general relativity quantized for small fluctuations around flat
space, is nonrenormalizable quantum field theory. In
\cite{Robinson:2005fj}, the authors used the background field method
to calculate the gravitational correction to the $\beta$ function
for the gauge theories. Their results should be reproduced in a
reliable fundamental theory of quantum gravity. Our investigation
provides the new insight into the weak gravity conjecture at the
quantum level and can be generalized to discuss the weak gravity
conjecture for the scalar field theories \cite{Huang:2007gk}.

\vspace{.5cm}

\noindent {\bf Acknowledgments}

We would like to thank P.M. Ho, S. Kim, K. Lee, M. Li, F.L. Lin, S.
Robinson, X. Wu and H. Yee for useful discussions. We also thank
department of physics in National Taiwan University for the
hospitality.

\end{document}